\documentclass[aip,prl,amsmath,amssymb,amstext,citeautoscript,punctuation,reprint]{revtex4-1}
\pdfoutput=1

\usepackage{hyperref}% include hyperlinks \href{text to display}{url}
\usepackage{graphicx,color,charter}% Include figure files
\usepackage{gensymb}

\begin{document}

\title{A Recommendation engine for suggesting unexpected thermoelectric chemistries
} %Title of paper



\author{Michael W. Gaultois}
%\homepage[URL: ]{www.michaelgaultois.com}
\email{mg757@cam.ac.uk}
\affiliation{Department of Chemistry, University of Cambridge, Cambridge, CB2 1EW, United Kingdom}

\author{Anton O. Oliynyk}
\affiliation{Department of Chemistry, University of Alberta, Edmonton, Alberta, T6G 2G2, Canada}

\author{Arthur Mar}
\affiliation{Department of Chemistry, University of Alberta, Edmonton, Alberta, T6G 2G2, Canada}

\author{Taylor D. Sparks}
\affiliation{Department of Materials Science and Engineering, University of Utah, Salt Lake City, Utah, 84112, USA}

\author{Gregory J. Mulholland}
\affiliation{Citrine Informatics, Redwood City, California, 94063, USA}

\author{Bryce Meredig}
\email{bryce@citrine.io}
\affiliation{Citrine Informatics, Redwood City, California, 94063, USA}

\date{\today}

 %Use showkeys class option if keyword display desired

\begin{abstract}
The experimental search for new thermoelectric materials remains largely confined to a limited set of successful chemical and structural families, such as chalcogenides, skutterudites, and Zintl phases.\cite{Nolas2001, Snyder2008NM, Nolas2006MB} In principle, computational tools such as density functional theory (DFT) offer the possibility of rationally guiding experimental synthesis efforts toward very different chemistries. However, in practice, predicting thermoelectric properties from first principles remains a challenging endeavor\cite{carrete2014finding}, and experimental researchers generally do not directly use computation to drive their own synthesis efforts. To bridge this practical gap between experimental needs and computational tools, we report an open machine learning-based recommendation engine (\href{http://thermoelectrics.citrination.com}{http://thermoelectrics.citrination.com}) for materials researchers that suggests promising new thermoelectric compositions, and evaluates the feasibility of user-designed compounds. We show that this engine can identify interesting chemistries very different from known thermoelectrics. Specifically, we describe the experimental characterization of one example set of compounds derived from our engine, $RE_{12}$Co$_5$Bi ($RE$\,=\,Gd, Er), which exhibits surprising thermoelectric performance given its unprecedentedly high loading with metallic $d$ and $f$ block elements, and warrants further investigation as a new thermoelectric material platform. We show that our engine predicts this family of materials to have low thermal and high electrical conductivities, but modest Seebeck coefficient, all of which are confirmed experimentally. We note that the engine also predicts materials that may simultaneously optimize all three properties entering into $zT$; we selected $RE_{12}$Co$_5$Bi for this study due to its interesting chemical composition and known facile synthesis.
\end{abstract}

\maketitle %\maketitle must follow title, authors, abstract and \pacs

\section{Introduction} \label{sec:intro}

For any materials problem, breaking out of ``local optima'' in composition space to discover entirely new chemistries remains a notoriously difficult challenge.\cite{jain2013commentary} Many of the most notable materials classes under investigation today--from Na$_x$CoO$_2$ derived thermoelectrics~\cite{Terasaki1997PRB} to iron arsenide superconductors~\cite{Kamihara2008JACS}--were discovered fortuitously. As a result, experimental efforts often gravitate toward incrementally improving \textit{known} chemistries (via doping, nanostructuring, etc.), as these efforts are more likely to bear fruit than high-risk searches through chemical whitespace for entirely new materials.

\begin{figure*}
\centering
\includegraphics[width=1\textwidth]{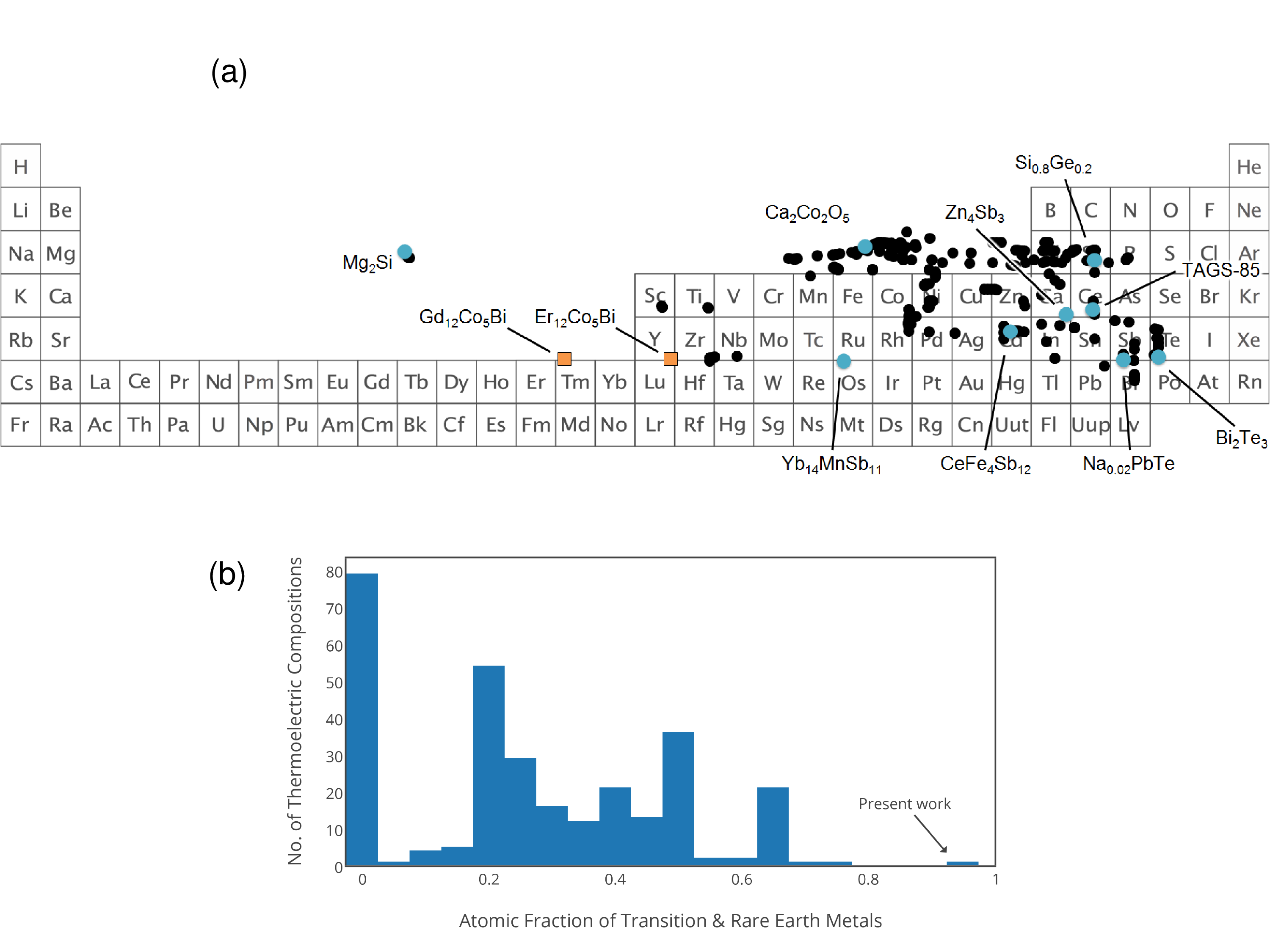}
\caption{(a) Most known thermoelectric materials lie in a tight cluster in composition space (black and blue dots; blue dots have chemical formulae explicitly labelled). The recommendation engine presented here allows the identification of new thermoelectric materials families that are well outside the existing composition space of common systems; in particular, we report the characterization of $RE_{12}$Co$_5$Bi ($RE$\,=\,Gd, Er; orange squares) in this work, which are chemically and structurally distinct from known thermoelectrics. (b) The strongly intermetallic $RE_{12}$Co$_5$Bi compounds we report here lie far outside the norm for metal loading among collected thermoelectric compositions in the Gaultois \textit{et al.}\cite{Gaultois2013CM} database.}
\label{fig:clustering}
\end{figure*}

The consequence of research communities' focus on further exploitation of known chemistries rather than exploration of unknown chemistries is that much of composition space simply remains uncharacterized. In Fig.\,\ref{fig:clustering}a, we illustrate the remarkable chemical homogeneity of most thermoelectric materials investigated to date. We plot each material from the thermoelectric database of Gaultois \textit{et al.}\cite{Gaultois2013CM} on the periodic table based on the composition-weighted average of the positions of elements in the material. The tight cluster of previously investigated chemistries is, as expected, dominated by chalcogenides and \textit{p}-block elements such as Sn and Sb. In contrast, we also show the positions of Gd$_{12}$Co$_5$Bi and Er$_{12}$Co$_5$Bi, materials derived from our recommendation engine, which we characterize as a new class of thermoelectrics in this work. These materials are almost pure intermetallics, in sharp contrast to thermoelectric compounds investigated to date (Fig.\,\ref{fig:clustering}b). The objective of our recommendation engine is to directly enable experimental researchers to rapidly identify new materials, such as $RE_{12}$Co$_5$Bi, that are very distinct from known compound classes, and worthy of further study.

\subsection*{A materials recommendation engine} \label{sec:recommendation_engine}

%Figure: Thermoelectric materials compositional phase space

Our recommendation engine is a machine learning-based approach\cite{meredig2014combinatorial, meredig2014dissolving} for efficiently driving synthetic efforts toward promising new chemistries. We have trained a machine learning model to make a confidence level prediction of whether the (1) Seebeck coefficient, (2) electrical resistivity, (3) thermal conductivity, and (4) band gap of input materials are within acceptable ranges for thermoelectric applications. We define these ranges as follows: (1) $|S|$\,$>$\,100\,$\mu$V\,K$^{-1}$; (2)\,$\rho$\,$<$\,$10^{-2}$\,$\Omega$\,cm; (3)\,$\kappa$\,$<$\,10\,W\,m$^{-1}$\,K$^{-1}$; and (4)\,$E_{\text{g}}$\,$>$\,0\,eV, all at room temperature. For each range, the engine gives a confidence score between 0 and 100\% that a given material's measured value for that property at room temperature will fall within the targeted range.
We would classify any material for which the answer to all these questions is likely ``yes'' as a potentially promising thermoelectric that may warrant further study. The purpose of our recommendation engine is thus neither to make \textit{quantitative} predictions of these thermoelectric properties, nor to definitively identify record-setting compounds--these remain open challenges for future work. Rather, the engine is intended to greatly augment the chemical intuition of experimental researchers working on materials discovery. In particular, we have found that our model's ability to screen vast numbers of possible compositions and short-list interesting candidates can inspire materials syntheses that would not have been obvious \textit{a priori}.

\section{Methods} \label{sec:methods}
\subsection*{Modelling and informatics}

Here we describe the approach used to construct the recommendation engine. Our engine is an example of materials informatics\cite{liu2006linking,rajan2005materials}, or the application of empirical machine learning methods to the prediction of materials behavior. Any machine learning approach for materials relies on three key ingredients: training data, descriptors, and choice of algorithm. Training data are the example sets from which the machine learning approach should extract meaningful chemical trends. Descriptors are the low-level characteristics of materials (e.g., crystal structure, chemical formula, etc.) that might correlate with materials properties of interest. Specifically, descriptors are either numerical (e.g., average atomic number $Z$) or categorial (e.g., crystal structure=perovskite) variables that enable us to ``vectorize'' materials in such a way that they become amenable to machine learning techniques. Finally, learning algorithms interrogate descriptor-vectorized training data for relevant patterns.

In this work, the training set comprises a large body of both experimental thermoelectric characterization data \cite{Gaultois2013CM} and first principles-derived electronic structure data \cite{jain2013commentary, ong2013python}. These data are publicly available via the Citrination platform (\href{http://www.citrination.com}{http://www.citrination.com}) and/or the Materials Project API (\href{http://www.materialsproject.org/open}{http://www.materialsproject.org/open}). These data consist of the Seebeck coefficients, thermal conductivities, electrical conductivities, and band gaps measured for thousands of materials as a function of temperature and a variety of other metadata conditions. Our model uses these input data to learn interesting chemical trends that could be exploited to design new materials. As large, high-quality training data sets are scarce in materials science relative to the biological sciences, where bioinformatics has become a standard tool, we urge the materials community to consider contributing to data infrastructures (Citrination, Materials Project, NIST's DSpace repository, EU's NoMaD, and others) that together will significantly expand open access to data for materials researchers.

Descriptors are the second key ingredient in materials informatics. The scientific literature around designing descriptors for materials has grown substantially in just the past several years\cite{curtarolo2013high, ghiringhelli2015big}. Indeed, recent work has shown that the predictive power of machine learning models for materials is strongly dependent upon the selected descriptor set.\cite{hansen2015machine} Our engine relies upon a tuned blend of descriptors designed in-house and drawn from a variety of sources.\cite{meredig2014combinatorial,carrete2014finding} By way of example, as materials scientists, we recognize that the periodic table contains a tremendous amount of information about how the elements behave and interact. We thus pre-bias our machine learning models with such knowledge (e.g., the $d$ block of the periodic table is metallic; Li and Na are chemically very similar but not identical; and the lanthanides behave similarly in ionic compounds). This step allows us to create predictive models with data sets that have thousands (rather than tens or hundreds of thousands) of examples.

Finally, our recommendation engine is built using the so-called random forest algorithm.\cite{breiman2001random} This algorithm constructs a large number of decision trees, all trained on slightly different subsets of the training data. Random forest is an ensembling technique, which takes advantage of the fact that a collection of ``weak'' learners such as decision trees can, in concert, model extraordinarily complex nonlinear behavior. An example rule that a single decision tree might learn is that if a material contains two elements with very different electronegativities (e.g., Na and Cl), that material is likely to have a large band gap. Of course, the thermoelectric phenomena we seek to model here are substantially more subtle, and thus a large random forest of decision trees is useful in untangling the underlying physics. We refer the reader elsewhere\cite{meredig2014combinatorial,carrete2014finding,blog} for more detailed discussions and tutorials on how to apply random forests to materials data.

\subsection*{Model validation} \label{sec:validation}
\begin{figure*}
\centering
\includegraphics[width=0.8\textwidth]{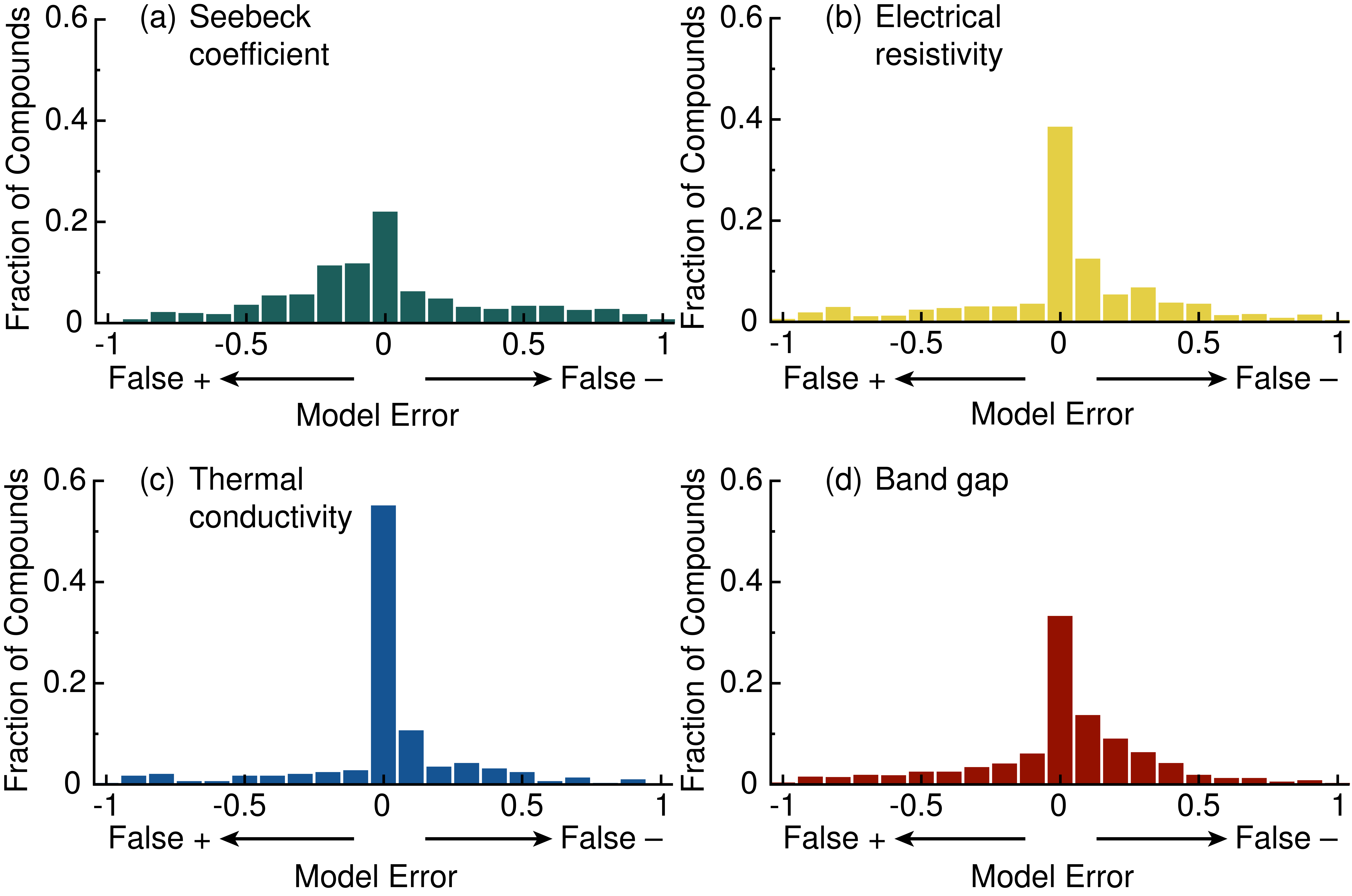}
\caption{Leave-one-out cross validation error histograms for the four key properties estimated by our recommendation engine: (a) Seebeck coefficient; (b) electrical resistivity; (c) thermal conductivity; and (d) band gap. For each material in our training set and each property, the recommendation engine gives a confidence score between 0 and 1 that the property value falls within the ideal windows we have defined for thermoelectric applications. Errors approaching $+1$ represent false negatives (our engine was extremely confident the material would be poor for that property, but the property is actually good); and an error of $-1$ is a false positive (our engine was extremely confident the material would be good for that property, but the property is actually poor). The peak around 0 for each property shows that the engine generally gives confidence values very close to unity for materials possessing properties in the desired ranges, or close to zero for materials whose property values fall outside the target range.}
\label{fig:model_validation}
\end{figure*}

We visualize the accuracy of our recommendation engine's predictions in Fig.\,\ref{fig:model_validation}, which represents the results of leave-one-out cross-validation (LOOCV) on our training data (in the case of the band gap data, we performed LOOCV on a subset of the extremely large training set). In the LOOCV procedure, if we have $n$ total measurements of a particular property such as thermal conductivity, we train our machine learning model on $n-1$ of these values and predict the $n$th (left out) value. We perform one training step and prediction for each property value, and present the error distribution for all $n$ values in Fig. \ref{fig:model_validation}. The error distribution then provides us with a sense of how we may expect the model to perform on new materials of which we have no prior knowledge.

%Figure. Validation of the model

Fig.\,\ref{fig:model_validation} indicates that our engine generally makes very reliable assessments of thermoelectric materials properties. The modes of the error distributions are in each case close to 0. For each property, the engine's errors skew toward false negatives (resistivity, band gap, thermal conductivity) or false positives (Seebeck), which reflects the fact that the underlying training data do not contain equal fractions of positive and negative examples. Seebeck coefficients prove most difficult to assess (\textit{i.e.}, the error distribution for that property has the largest standard deviation), likely because there are strikingly different mechanisms that underpin the values, for example, strongly correlated oxides as opposed to degenerate semiconductors. Owing to the difficulty in assessing the Seebeck coefficient, initial predictive models using only the electrical resistivity, thermal conductivity, and Seebeck coefficient produced too many candidates that were good metals with poor Seebeck coefficients. To remedy this shortcoming and provide more robust recommendations, the band gap was added as a secondary metric, where we determine the probability whether a given composition will have a non-zero bandgap.

\subsection*{Experimental details} \label{sec:experimental}

$RE_{12}$Co$_5$Bi ($RE$\,=\,Gd, Er) samples were made by arc-melting freshly filed Er or Gd pieces (99.9\%, Hefa), Co powder (99.8\%, Cerac), and Bi powder (99.999\%, Alfa Aesar). Stoichiometric mixtures (0.5\,g total mass) with 5-7\% excess of bismuth were pressed into pellets and melted twice in arc-melting furnace under argon atmosphere (Edmund B\"{u}hler Compact Arc Melter MAM-1). The total mass loss after melting was $<$\,1\%. The samples were sealed in silica tubes and annealed at 1070\,K for one week, then quenched in cold water. To produce enough material for physical property measurement, $\sim$70 samples of each compound were prepared, and pure samples were combined by melting into a single ingot of $\sim$5\,g, which was sanded to yield the appropriate geometry (either a rectangular bar, or a cylinder). Density was measured using Archimedes' method; the final pellets had densities 100\% of the single crystal values ($\rho_{\text{Gd}_{12}\text{Co}_5\text{Bi}}$\,=\,8.6\,g/cm$^3$, $\rho_{\text{Er}_{12}\text{Co}_5\text{Bi}}$\,=\,9.9\,g/cm$^3$). 

Powder X-ray diffraction patterns were collected using an INEL CPS 120 diffractometer with Cu\,K$\alpha_1$ radiation at room temperature, and Rietveld refinement was used to confirm the structure and phase purity (see Supporting Information). Backscatter electron microscopy and elemental analysis via energy dispersive X-ray spectroscopy (EDX) were performed with a JEOL JSM-6010LA InTouchScope scanning electron microscope. Backscatter microscographs reveal the samples are largely compositionally homogeneous (see Supporting Information). Quantitative elemental analysis on several polished pieces found an atomic composition of Gd69(2)Co26(2)Bi5(2) which is in a good agreement with expected $RE_{12}$Co$_5$Bi composition. Er$_{12}$Co$_5$Bi samples were not appropriate for quantitative analysis because of overlaping Co K$\alpha$ (6.924\,keV) and Er L$\alpha$ (6.947\,keV) lines.

High-temperature thermoelectric properties (electrical resistivity and Seebeck coefficient) were measured with an ULVAC Technologies ZEM-3. Sample bars had approximate dimensions of 9\,mm$\times$4\,mm$\times$4\,mm. Measurements were performed with a helium under-pressure, and data was collected from 300\,K to 800\,K through three heating and cooling cycles over 18 hours to ensure sample stability and reproducibility.

\section{Discussion} \label{sec:discussion}
In this work, we are interested not only in developing a model that gives accurate predictions of materials properties, but also in making it immediately accessible and useful for experimental researchers. To that end, we have published our recommendation engine as a web app at \href{http://thermoelectrics.citrination.com}{http://thermoelectrics.citrination.com}, where researchers may explore a pre-computed list of around 25,000 known compounds (representing a sizable subset of the Inorganic Crystal Structure Database, or ICSD), and also use our model to evaluate in real-time their own materials candidates. In this way, we hope that the app serves as a rapid triage tool for ideas for potential new thermoelectric materials. Our pre-computed list may be arranged according to the probabilities associated with any one of the four properties we are modelling, and is sorted by default according a composite score that takes all four properties into account. Furthermore, the user may specify cutoff thresholds for any of the properties, and thereby greatly reduce the size of the list.

As we believe our extensive precomputed list contains some interesting and heretofore uncharacterized candidate thermoelectric materials, we now comment on a select set of high-ranking compounds. Several of these compounds are given in Table \ref{table:topranked}.

%Table: example materials
\begin{table*}
\centering

\caption{Several promising new thermoelectric compounds selected from our pre-computed list. The $P$ values refer to the engine's confidence level that a given material will exhibit a room-temperature value for a particular property (e.g., $S$ or $\rho$) within the target ranges specified above. The full compound list is available for exploration at \href{http://thermoelectrics.citrination.com}{http://thermoelectrics.citrination.com}.}
\label{table:topranked}
\begin{tabular}{l c c c c c l} % centered columns (4 columns)
\hline\hline %inserts double horizontal lines
Material & P$_{S}$ & P$_{\rho}$ & P$_{\kappa}$ & P$_{gap}$ & Composite & Comments \\ [0.5ex] % inserts table 
%heading
\hline % inserts single horizontal line
TaPO$_5$ and TaVO$_5$ & 0.894 & 0.793 & 0.958 & 0.987 & 3.537 & High polyhedral connectivity and structural superlattices \\ % inserting body of the table
Tl$_9$SbTe$_6$ & 0.845 & 0.871 & 0.999 & 0.876 & 3.46 &Recently reported to be a good thermoelectric (zT\,$\approx$\,1 at 600\,K) \\
TaAlO$_4$ & 0.893 & 0.703 & 1 & 0.977 & 3.477 & High mass contrast, high polyhedral connectivity \\&&&&&& (edge- and corner-sharing TaO$_6$ octahedra) \\
SrCrO$_3$ & 0.772 & 0.767 & 0.996 & 0.95 & 3.308 & High polyhedral connectivity (3-D corner-sharing CrO$_6$ octahedra), \\&&&&&& metallic when made under high pressure \\
TaSbO$_4$ & 0.892 & 0.919 & 1 & 0.997 & 3.559 & High polyhedral connectivity: layered, edge-sharing MO$_6$ octahedra \\ [1ex] % [1ex] adds vertical space
\hline %inserts single line
\end{tabular}
\end{table*}

TaVO$_5$ and TaPO$_5$ occur in an analogous crystal structure to the phosphate tungsten bronzes \cite{Chahboun1986JSSC,Chahboun1988MRB}. These materials  can be expected to have good thermoelectric performance given the heavy atoms, the potential for low electrical resistivity provided by the repeating ReO$_3$-type structural network that is highly connected in three dimensions, and the intrinsic crystallographic shear provided by the crystal structure. Although the phosphate tungsten bronzes themselves are not highly rated, their metallic electrical transport properties are encouraging for structural analogues \cite{Greenblatt1993IJMPB}. Moreover, TaVO$_5$ has a negative coefficient of thermal expansion and a structural transition at $600 \degree $C \cite{Wang2011IC}. This structural transition may lead to softening of phonon modes and anharmonic scattering, which may lead to low thermal conductivity. The second material of interest we present is Tl$_9$SbTe$_6$. Though this compound was not included in the thermoelectric database, it scores highly within the recommendation engine, and good thermoelectric performance has been subsequently demonstrated in recent work \cite{Guo2013CM}.

The suggestion of TaAlO$_4$, SrCrO$_3$, TaSbO$_4$ and other oxides expected to be insulators can be understood because the recommendation engine uses as training data references where stoichiometric formulas were primarily reported rather than doping details.\cite{Depero1997JSSC, Ok2001JSSC} Nevertheless, with doping through substitution or reduction, these compound may exhibit moderate electrical performance. Further, these materials all feature extended structures that are highly connected in three dimensions, an important feature for low electrical resistivity. Moreover, the large mass contrast on the cation sublattice in TaAlO$_4$ (edge shared TaO$_6$ and AlO$_6$ octahedra) could lead to low thermal conductivity, and previous reports have shown that SrCrO$_3$ is metallic when synthesized under pressure \cite{Komarek2011PRB}.

Many of the high-ranking candidate materials are interesting because of their highly connected extended structures, even though the recommendation engine does not use features of crystal structure to make its suggestions. The chief disadvantage to training prediction algorithms using crystal structure is that structure then becomes a \emph{required input} for making predictions, and yet structure is by definition not available for uncharacterized materials. However, the absence of crystal structure does cause our engine difficulty where changes in crystal structure with similar elemental compositions cause large changes in physical properties. For example, both DyPO$_4$ and LaPO$_4$ are predicted to have low thermal conductivity. However, LaPO$_4$ is monazite, a corner edge-shared structure, whereas DyPO$_4$ is xenotime \cite{Ni1995AM}, an edge-shared structure leading to inherently higher thermal conductivity.\cite{Winter2007JACS}

\subsection*{New materials and their properties} \label{sec:re12co5bi}

\begin{figure*}
\centering
\includegraphics[width=0.5\textwidth]{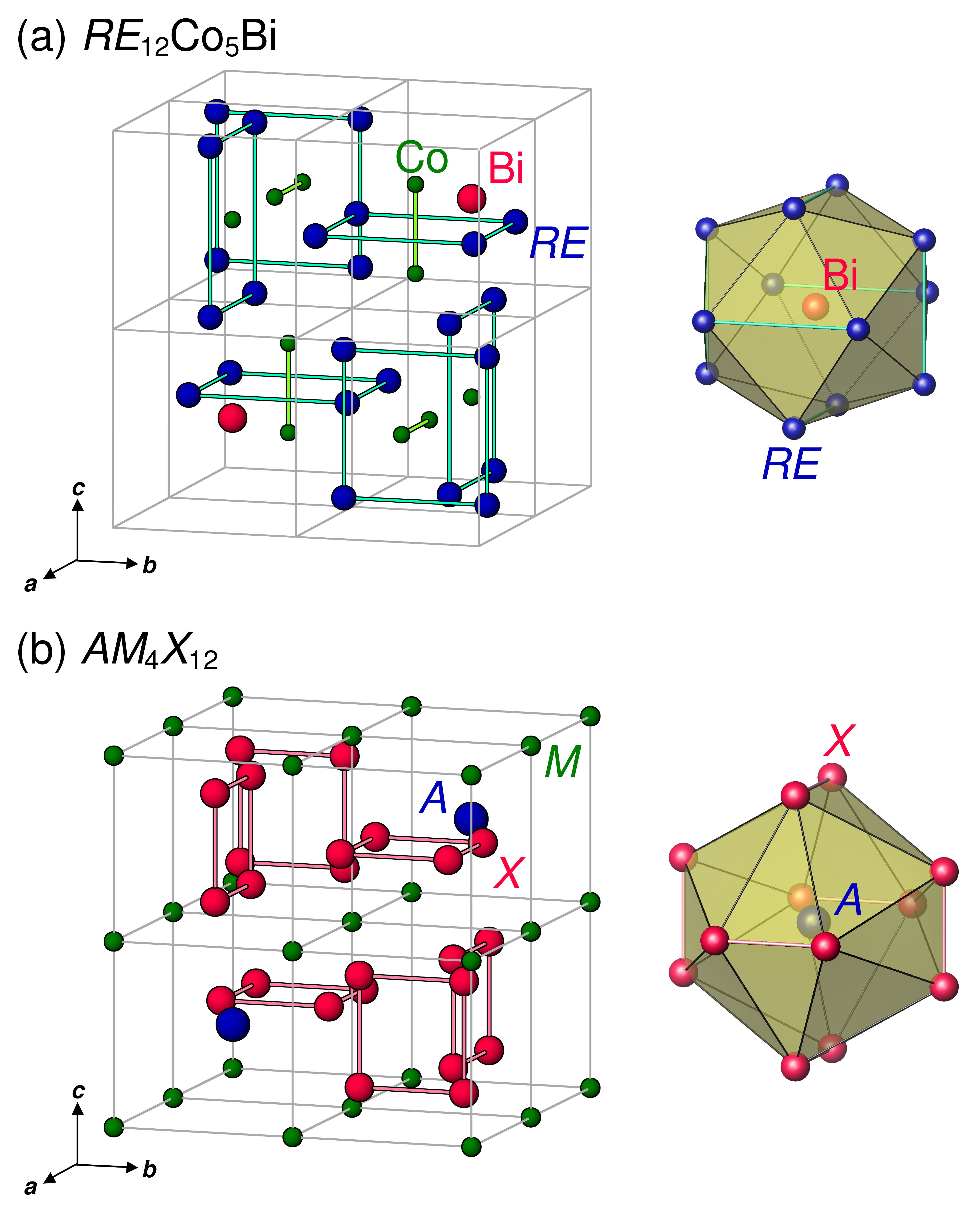}
\caption{(a) Crystal structure of $RE_{12}$Co$_5$Bi (prototype Ho$_{12}$Co$_5$Bi), of which Er$_{12}$Co$_5$Bi and Gd$_{12}$Co$_5$Bi are exemplars. (b) Crystal structure of the filled skutterudites, which have the generic chemical formula $AM_{4}X_{12}$. These two structure types share an icosahedral motif consisting of $RE_{12}$Bi and $AX_{12}$ units, respectively.}
\label{fig:structure}
\end{figure*}

Our final and most important task in this work is to demonstrate that our recommendation engine can indeed guide researchers toward interesting experimental discoveries. Among the set of high-scoring candidate materials, we selected Er$_{12}$Co$_5$Bi and Gd$_{12}$Co$_5$Bi to characterize as thermoelectric materials due to their facile synthesis through arc melting, and due to the fact they are chemically quite distinct from known thermoelectrics (Fig. \ref{fig:clustering}). While the $RE_{12}$Co$_5$Bi ($RE$\,= rare earth) family of compounds has only been sparsely studied in the literature, their crystal structure and initial low-temperature electrical and magnetic properties have been reported by Mar and coworkers~\cite{Tkachuk2005IC}. The crystal structure of $RE_{12}$Co$_5$Bi is shown in Figure\,\ref{fig:structure}. 

%Figure: Crystal structure 

Interestingly, the crystal structure of our candidate thermoelectric exhibits notable similarity to the structures of known thermoelectrics, in spite of the fact that crystal structure was not an input feature for our recommendation engine. Ho$_{12}$Co$_5$Bi is the eponymous structure prototype (orthorhombic, space group \textit{Immm}) adopted by a series of rare-earth intermetallics $RE_{12}$Co$_5$Bi ($RE$\,=\,Y,\,Gd,\,\ldots,\,Tm). In this structure, the Ho$_{12}$Bi icosahedra play an analogous role to the LaP$_{12}$ icosahedra in the filled skutterudite prototype LaFe$_4$P$_{12}$; rare-earth atoms ``rattling'' within their 12-fold coordinated cages is the idiosyncratic feature of filled skutterudites that imparts low thermal conductivity so prized in thermoelectric materials. In fact, if the transition metal atoms, which occupy different sites in these structures, are disregarded, the Ho$_{12}$Bi framework is an antitype to the LaP$_{12}$ framework, with the roles of the rare-earth and group 15 elements reversed. We hypothesize its crystallographic similarity to skutterudite could be partly responsible for the thermoelectric behavior of $RE_{12}$Co$_5$Bi ($RE$\,=\,Gd, Er).

%Figure: Thermoelectric properties 
\begin{figure*}
\centering
\includegraphics[width=0.7\textwidth]{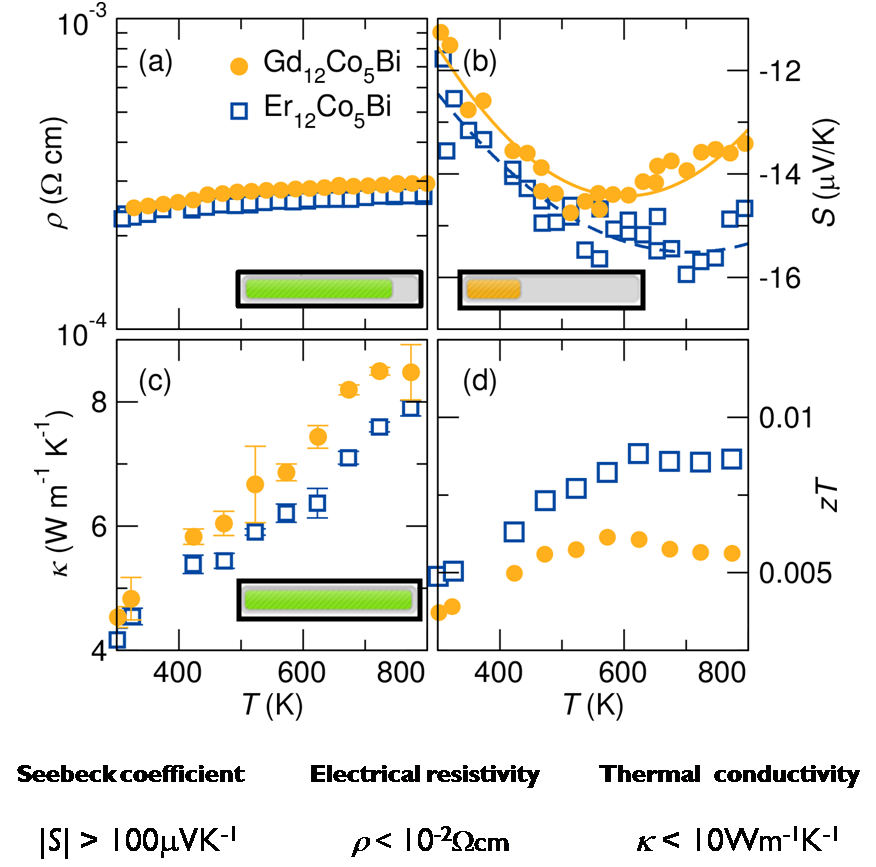}
\caption{Thermoelectric characterization of $RE_{12}$Co$_5$Bi ($RE$\,=\,Gd, Er). (a) Electrical resistivity, (b) Seebeck coefficient, (c) thermal conductivity, and (d) thermoelectric figure of merit $zT$ as a function of temperature. We also include the recommendation engine's confidence levels for the first three properties; the lowest-probability property, the Seebeck coefficient, is indeed found to be below the 100$\mu$V/K threshold.}
\label{fig:TE_properties}
\end{figure*}

We give a full thermoelectric characterization of Er$_{12}$Co$_5$Bi and Gd$_{12}$Co$_5$Bi in Fig.\,\ref{fig:TE_properties}. Based on these results, we report the discovery of a new thermoelectric class, which remains a completely unoptimized, pure bulk material and thus lends itself to further study. Notably, the material falls far outside the usual search space for thermoelectrics (Fig. \ref{fig:clustering}), and was neither the result of simple interpolation between known compounds nor obvious from a strict chemical intuition standpoint. The electrical resistivity is commensurate with other high-performing materials such as chalcogenides, although the Seebeck coefficient is too low for the material to be competitive with the best-known thermoelectrics. Furthermore, the thermal conductivity is relatively high, but the filled cage structure lends itself to substitution that has successfully reduced thermal conductivity in the skutterudite systems \cite{Nolas1996JAP,Nolas2006MB}. In $RE_{12}$Co$_5$Bi ($RE$\,=\,Gd, Er), the thermal conductivity from 300\,K to 800\,K ranges from 4\,W\,m$^{-1}$\,K$^{-1}$ to 8\,W\,m$^{-1}$\,K$^{-1}$, comparable to the half-Heuslers \cite{Graf2011PSSC,Douglas2012APL}. Note that, as illustrated in Fig. \ref{fig:TE_properties}, these results are in accord with the engine's predictions; the models give a high probability of achieving the thresholds for electrical conductivity (a) and thermal conductivity (c) (see confidence bar insets), while also suggesting a low probability of observing a large Seebeck coefficient (b). The electrical performance figure of merit $\kappa zT$ is around 0.03\,W\,m$^{-1}$\,K$^{-1}$ at 400 K, which is actually higher than that of nearly 30\% of the thermoelectrics in the Gaultois \textit{et al.} thermoelectrics database;\cite{Gaultois2013CM} of course, the database is a highly self-selected set of materials, consisting of literature-reported thermoelectrics, and would skew toward much higher $\kappa zT$ values than would a random subset of all crystalline materials. We note, of course, that the \textit{zT} of several other thermoelectric materials can be significantly improved through carrier concentration tuning and microstructural engineering. For example, undoped polycrystalline Si has a 60-fold increase in performance after optimization, going from $zT<0.01$ to 0.6 at 300\,K~\cite{Hochbaum2008N}. 
%nanoporous Ge http://scitation.aip.org/content/aip/journal/apl/95/1/10.1063/1.3159813#c5
%Si nanowires http://www.nature.com/nature/journal/v451/n7175/abs/nature06381.html
%pure Si http://journals.aps.org/pr/abstract/10.1103/PhysRev.167.765

Another observation from Fig. \ref{fig:TE_properties} illustrates the scientific boon of studying entirely new classes of materials. Unexpectedly, $RE_{12}$Co$_5$Bi ($RE$\,=\,Gd, Er) exhibits \textit{increasing} thermal conductivity with temperature. (We note the recommendation engine successfully chose a material with a low thermal conductivity at room temperature, which would normally decrease with increasing temperature.) The increasing electrical resistivity with temperature indicates metallic electrical transport, so the electrical contribution to the total thermal conductivity should therefore decrease with increasing temperature. Additionally, the phonon contribution to thermal conductivity should also decrease with increasing temperature due to more phonon--phonon (Umklapp) scattering \cite{Grimvall1999}. Thermal conductivity is calculated from the following relation: $\kappa$\,=\,$\alpha$\,$\rho$\,$C_{\text{p}}$, where $\alpha$ is thermal diffusivity, $C_{\text{p}}$ is heat capacity, and $\rho$ is density. Normally, thermal diffusivity has a negative temperature dependence whereas heat capacity and density both have positive temperature dependence. However, for this compound we observe a \textit{positive} temperature dependence for the thermal diffusivity even after multiple measurements, the origin of which is not presently understood. Materials with increasing thermal conductivity with temperature are rare, though not unprecedented \cite{Jacobsson1968PRB,Tritt2004}, and further studies on this class of compounds to shed light on this anomaly could thus lead to new strategies for thermoelectric materials optimization.

\section{Conclusions} \label{sec:conclusions}
This initial experimental validation of our recommendation engine is encouraging. The present work represents the first time that machine learning has been used to suggest an experimentally viable new compound from true chemical white space, where no prior characterization had hinted at promising chemistries. The implication is that our approach--wherein a data-driven computational tool directly augments experimental capabilities and intuition--is a semi-rational way to discover new materials families that may have desirable properties. We suggest that such an paradigm could eventually replace trial-and-error and fortuity in the search for new materials across a wide variety of application areas.

\section*{Acknowledgments} \label{sec:ack}
We thank Ram Seshadri for helpful discussions and insight. We thank the National Science Foundation for support of this research through NSF-DMR 1121053, as well as the Natural Sciences and Engineering Research Council of Canada (NSERC). Additionally, this research made extensive use of shared experimental facilities of the Materials Research Laboratory: a NSF MRSEC, supported by NSF-DMR 1121053. MWG thanks the NSERC for support through a Postgraduate Scholarship, and the US Department of State for support through an International Fulbright Science \& Technology Award. BM and GJM are founders and significant shareholders in Citrine Informatics Inc.

%\end{acknowledgments}

\bibliography{recengine}

\end{document}